\documentclass[twocolumn]{revtex4-1}
\usepackage[utf8]{inputenc}

\usepackage{amssymb}
\usepackage{amsmath,amscd}
\usepackage{graphicx}
\usepackage{dcolumn}

\usepackage{color}
\usepackage{hyperref}
\usepackage{natbib}

\newcommand{\Z}{{\mathcal Z}}

\begin{document}

\title{Antiferromagnetic Ising model in an imaginary magnetic field}
\author{Vicente Azcoiti}
\affiliation{Departamento de Física Teórica, Facultad de Ciencias,
  Universidad de Zaragoza \\ C/Pedro Cerbuna 12, E-50009, Zaragoza
  (Spain)}
\author{Giuseppe \surname{Di Carlo}}
\affiliation{INFN, Laboratori Nazionali del Gran Sasso, \\
I-67100 Assergi, L'Aquila (Italy)}
\author{Eduardo Follana}
\affiliation{Departamento de Física Teórica, Facultad de Ciencias,
  Universidad de Zaragoza \\ C/Pedro Cerbuna 12, E-50009, Zaragoza
  (Spain)}
\author{Eduardo Royo-Amondarain}
\email{Corresponding author: eduroyo@unizar.es}
\affiliation{Departamento de Física Teórica, Facultad de Ciencias,
  Universidad de Zaragoza \\ C/Pedro Cerbuna 12, E-50009, Zaragoza 
  (Spain)}

\begin{abstract}

We study the two-dimensional antiferromagnetic Ising model with a
purely imaginary magnetic field, which can be thought of as a toy
model for the usual $\theta$ physics. Our motivation is to have a
benchmark calculation in a system which suffers from a strong sign
problem, so that our results can be used to test Monte Carlo methods
developed to tackle such problems.

We analyze here this model by means of analytical techniques,
computing exactly the first eight cumulants of the expansion of the
effective Hamiltonian in powers of the inverse temperature, and
calculating physical observables for a large number of degrees of
freedom with the help of standard multi-precision algorithms. We
report accurate results for the free energy density, internal energy,
standard and staggered magnetization, and the position and nature of
the critical line, which confirm the mean-field qualitative picture,
and which should be quantitatively reliable, at least in the
high-temperature regime, including the entire critical line.

\end{abstract}

\maketitle

\section{Introduction}

One of the major challenges for high-energy and solid-state theorists
is the numerical simulation of systems with a severe sign problem.
If we denote the microscopic states of a given
physical system by $s$, and the thermodynamics of such system is
described by a partition function of the form $\Z = \sum_{s} P(s)$, we
say that the system in question presents a sign problem if the
``weights'' $P(s)$ are not real and positive: This implies that we
cannot interpret $P(s)$ as a proper probability distribution, and the
standard, efficient Monte Carlo algorithms cannot be applied. Not all
sign problems are equally severe. Let us restrict ourselves for
simplicity to the case where the $P(s)$ are real but not positive
definite\footnote{The discussion for complex weights does not add any
  fundamental difficulty.}. One can easily devise a reweighting
algorithm that uses the absolute value $\left|P(s)\right|$ as the
weight of each state, and shifts the sign of $P(s)$ into the
observables. Now a standard Monte Carlo method is applicable, and in
the limit of infinite statistics we should obtain the correct
result. With finite statistics, however, a key quantity is the
thermodynamic average of the sign of each contribution to the
partition function, that is, $\left< \text{sign}(P(s))\right>$. If
this quantity goes to zero exponentially with the volume, $\left<
\text{sign}\right> \propto e^{-\alpha V}$, then we would need an
exponential amount (in the volume of the system $V$) of statistics to
get correct results, which is of course impossible in practice. In
this case we say that the sign problem is severe.

QCD at finite baryon density, QCD with a topological
term in the action, chains of quantum spins with antiferromagnetic
interactions, the two-dimensional O(3) non linear sigma model with a
topological term, and the Hubbard model are some of the most popular
examples of relevant physical systems where a SSP is present. The
existence of a SSP is the main reason for the little progress made on
the theoretical understanding of these physical systems outside of
phenomenological models.

In order to check novel Monte Carlo methods designed to tackle such
problems, it is highly desirable to have a set of benchmark
calculations as extensive as possible. For very few systems an
analytic solution is known, for example, the one-dimensional
antiferromagnetic Ising model with an imaginary magnetic field, the
two-dimensional compact U(1) model with topological term, or the
two-dimensional Ising model with an imaginary magnetic field $h = i
\pi/2$. In a few other cases the sign problem can be avoided by
reformulating the physical system with new degrees of freedom, taking
advantage of the fact that a good choice of these degrees of freedom
provides an equivalent physical system free from the sign problem,
which can therefore be simulated by standard
methods.\footnote{Unfortunately this idea works only in a few cases
  which, until now, are not the most interesting physical
  systems. Indeed none of the examples previously mentioned have been
  solved with this idea.}

Our motivation for this paper is to provide a benchmark calculation for
a system for which we do not have an analytic solution available, nor
a reformulation that avoids the sign problem. We study the 
two-dimensional antiferromagnetic Ising model with a purely imaginary
magnetic field, which can be thought of as a toy model for the usual
$\theta$ physics. Indeed the Euclidean partition function for QCD with
a nonvanishing $\theta$ term can be written in the form
\begin{equation}
\Z_V(\theta) = \sum_n p_V(n) e^{i \theta n}
\end{equation}
where $n$, the topological charge, is an integer, and $p_V(n)$ is, up
to a normalization, the probability of the topological sector $n$ at
$\theta = 0$. This has the same structure as the partition function of
the antiferromagnetic Ising model in an external purely imaginary
magnetic field, as we will see in detail later on, and we expect that
the SSP in both systems should also be similar. 

This system was studied in \cite{Shrock:2008} by locating the zeros
of the partition function in the complex temperature-magnetic field
plane, and they find, for purely imaginary magnetic field, a rich
phase structure with two phases characterized by a vanishing
(nonvanishing) staggered magnetization, separated by a phase
transition line. We study this system by an exact cumulant expansion
to eighth order, followed by the analytic computation of the partition
function and other physical quantities for a large number of degrees
of freedom with the help of a standard multiprecision algorithm. This
amounts essentially to the computation of the effective Hamiltonian up
to order $T^{-8}$, and therefore is expected to work well in the 
high-temperature regime, and we provide strong evidence that this is
indeed the case. Our results are consistent with \cite{Shrock:2008},
and extend the results of \cite{Azcoiti:2011ei}, obtained through the
application of algorithms developed in
\cite{Azcoiti:2002vk,Azcoiti:2003vv}, and through a mean-field
analysis. We are able to obtain a more precise quantitative
determination of the transition line separating the paramagnetic and
antiferromagnetic phases of the model.

For some systems with a SSP, we know \textit{a priori} that the partition
function will be positive, for example systems in thermal equilibrium
with a (Hermitian) Hamiltonian description. Such is the case in a
quantum field theory with a $\theta$ term. In the toy model we study
here, although we do not have a rigorous proof in this
case,\footnote{This would imply a nontrivial restriction on the
  position of the Lee-Yang zeros for the antiferromagnetic Ising
  model. To the best of our knowledge, very little is rigorously known
  about such zeros.} we have evidence that, at least in the region
where the approximation we use is valid, the partition function is
indeed positive (it is trivially always real).

Such evidence is twofold. First, we can prove rigorously that up to
the fifth cumulant, the partition function is indeed
positive. Unfortunately we have not been able to extend this proof to
higher cumulants, but in our multiprecision calculations with up to eight
cumulants, we have never seen an instance where the partition function
is negative or vanishes. This is highly nontrivial: If instead of a
constant imaginary magnetic field we try, for example, to put a
staggered imaginary field in our lattice (this is of course equivalent
to the ferromagnetic model with a constant imaginary field), we
immediately get a fluctuating sign for the partition function.

Second, there have been studies locating the Lee-Yang zeros of the
two-dimensional antiferromagnetic Ising model up to $14^2$ lattices
\cite{Kim}, and in $12\times13$ lattices \cite{Shrock:2008}.  Up to
that size there is no sign of any zeros cutting the imaginary axis at
any temperature.

Whereas this by no means amounts to a rigorous proof, we believe it
provides a strong indication that, at least in the region of interest
for this paper, this model should have a positive partition
function.

This paper is organized as follows. Section \ref{2dmodel} is devoted
to formulate the model and to recall the main ingredients and results
of the mean-field approximation developed in \cite{Azcoiti:2011ei}. In
Sec. \ref{cumulant_expansion} we introduce the cumulant expansion,
report the analytical results for the first eight cumulants in the
two-dimensional model, and write the analytical expressions for the
free energy and mean values of interesting physical quantities. The
results for the staggered magnetization, susceptibility, and phase
diagram of the model are reported in Sec. \ref{results}, where we
also compare our results at $h = 0$ and $i \pi/2$ with the
analytical solutions of \cite{Onsager:1943jn,Lee:1952ig,McCoy:1967zz}.
In Sec. \ref{conclusions} we report our conclusions. The technical
details of the analytical computation of the cumulant expansion can be
found in Appendix \ref{appendixcumulants}, and several tables with
numerical results can be found in Appendix \ref{appendixtables}.

\section{Two-dimensional Ising model}
\label{2dmodel}

The Ising model \cite{Ising:1925em, Onsager:1943jn,
  Lee:1952ig,McCoy:1967zz, Matveev:1994ha,McCoy:1973,Shrock:2008} has
been studied for a long time now, and it has known analytical
solutions in the one-dimensional case at any external magnetic field
$h$ \cite{Ising:1925em}, and in two dimensions only for the case without
magnetic field $h$ \cite{Onsager:1943jn} and for $h = i\theta/2 = 
i\pi/2$ \cite{Lee:1952ig, McCoy:1967zz}. The model with a pure imaginary magnetic field
suffers from a SSP in any number of dimensions. In addition to that,
the expected phase diagram for $d\geq 2$ is non trivial
\cite{Azcoiti:2011ei}, making the reconstruction of the
$\theta$ dependence of the observables even more challenging. All this
makes the model a good theoretical laboratory to test new methods
designed to deal with the SSP. It is therefore worthwhile to carry out
a detailed study of this model at purely imaginary magnetic field,
particularly because little progress has been achieved on
reconstructing the $\theta$ dependence of the observables, apart from
the analysis of \cite{Azcoiti:2011ei} and the recent study in
\cite{DeForcrand}.

The partition function of the model, following the conventions of
\cite{Azcoiti:2011ei}, is:
\begin{equation}\label{Zpart1}
\Z = \sum_{\{s_i\}}\exp{\left(F\sum_{<ij>}{s_is_j}
+i\theta\dfrac{1}{2}\sum_{i}{s_i}\right)}.
\end{equation}
The half magnetization
\begin{equation}
\dfrac{M}{2}\equiv\dfrac{1}{2}\sum_i s_i,
\end{equation}
is an integer taking any value between $-N/2$ and $N/2$, where $N$ is
an even number denoting the total number of spins in the lattice. It
is in this sense that we identify $M/2$ with a topological charge and
regard the imaginary magnetic field term in the action as a
$\theta$ term. It is important to mention that, from now on, we will
consider only the antiferromagnetic case $F<0$, since the model with
imaginary field does not define a unitary theory for arbitrary values
of the ferromagnetic coupling \cite{Lee:1952ig,Azcoiti:1999rq}.

As we shall see in detail in the next section, by dividing the
rectangular lattice into two sublattices, introducing the respective
magnetizations $M_1$ and $M_2$, making a cumulant expansion and
keeping only the first cumulant, we arrive at the following
approximation to the partition function (where $d$ denotes the
dimensionality of the lattice):
\begin{equation}\label{Z1c}
\Z_{1c}(F,\theta) = \sum_{\{s_i\}}\exp \left(i\theta\dfrac{M_1+M_2}{2}
+4\dfrac{Fd}{N}M_1M_2 \right).
\end{equation}
We recall now the mean-field analysis carried out in
\cite{Azcoiti:2011ei}. The resulting partition function,
\begin{equation}\label{Zmf}
\Z_{MF}(F,\theta) = \sum_{\{s_i\}}\exp\left(i\theta\dfrac{M_1+M_2}{2}
- \dfrac{Fd}{N}(M_1-M_2)^2\right),
\end{equation}
is different from Eq. (\ref{Z1c}). However, it can be seen to give the
same qualitative results for the observables and the phase diagram. In
this regard, we will consider the first-cumulant expansion $\Z_{1c}$
as a mean-field approximation to $\Z$, and the general expansion
itself as an improvement of it, at least for small $F$, where the
expansion is expected to converge.

\begin{figure}
        \includegraphics[scale=1]{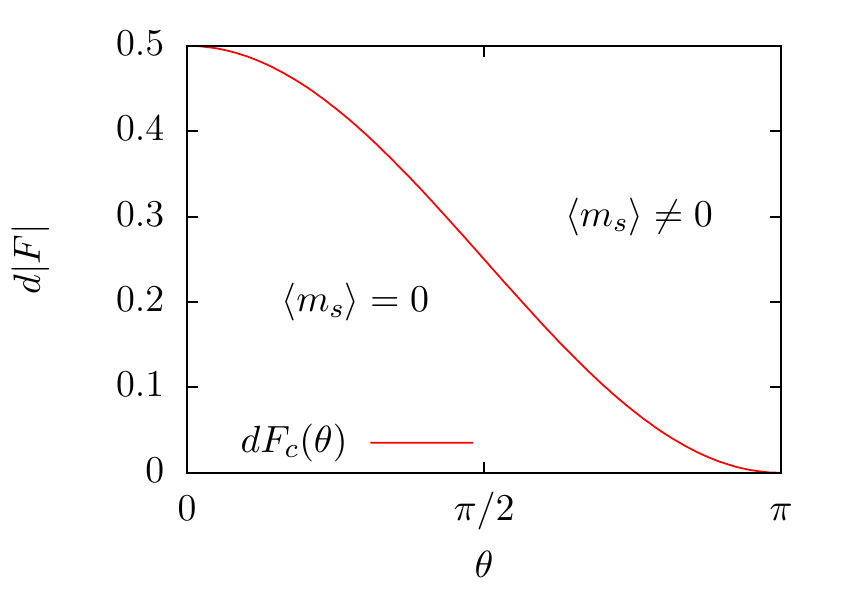}
    \caption{Phase diagram of the mean-field approach of 
\cite{Azcoiti:2011ei} to the antiferromagnetic Ising model in the 
$F-\theta$ plane.}\label{phDiag}
\end{figure}

Applying standard saddle-point techniques to the mean-field partition
function \cite{Azcoiti:2011ei}, one obtains the $F-\theta$ phase diagram
shown in Fig. \ref{phDiag}. A second order critical line,
\begin{equation}
dF_c=\dfrac{1}{2}\cos^2{\dfrac{\theta_c}{2}},
\end{equation}
separates two different phases: a staggered one, with $\langle
m_s\rangle\neq 0$, for $\theta\geq \theta_c$, and a paramagnetic one,
with $\langle m_s\rangle=0$, for $\theta<\theta_c$.

\section{Cumulant expansion and observables}
\label{cumulant_expansion}

Our interest is focused on the antiferromagnetic model, where the
staggered magnetization is a good order parameter. From now on we will
work with a rectangular two-dimensional lattice, although the method
is easily generalizable to any number of dimensions. We divide the
lattice into two sublattices $\Omega_1$ and $\Omega_2$ in a chessboard
fashion. In the two-dimensional lattice this means that if $i$ and $j$
index, respectively, the row and the column of a given spin, this spin
will be in the first (second) sublattice if the sum $i+j$ is even
(odd). For simplicity we will require that both lengths of the lattice
be even. Denoting by $N$ the total number of points in the lattice, we
define the magnetization densities $m_1$ and $m_2$ as
\begin{equation}
m_j\equiv\dfrac{M_j}{N/2}\equiv\dfrac{\sum_{i\in\Omega_j}s_i}{N/2}
\qquad j=1,2,
\end{equation}
and the density of staggered magnetization is
\begin{equation}
m_s \equiv \dfrac{m_1 - m_2}{2}.
\end{equation}

Let us denote by $g(m_1,m_2)$ the number of microstates with
magnetization densities $m_1$ and $m_2$ in sublattices $\Omega_1$ and
$\Omega_2$, respectively, that is,
\begin{equation}
g(m_1,m_2)=\sum_{\{s_i\}}\delta\left(\sum_{i\in\Omega_1}s_i-M_1\right)
\delta\left(\sum_{i\in\Omega_2}s_i-M_2\right).
\end{equation}
A trivial computation gives:
\begin{equation}
g(m_1,m_2)=\binom{N/2}{N_{1+}}\binom{N/2}{N_{2+}},
\end{equation}
with $N_{j+}\equiv N(1+m_j)/4$ for $j=1,2$.
Defining now the expected value at fixed $m1,m2$ as:
\begin{eqnarray}
\left<\mathcal{O}\right>_{m_1,m_2} &\equiv & \frac{1}{g(m_1,m_2)}         
\nonumber \\*
& &\times \sum_{\{s_i\}}\delta(\sum_{i\in\Omega_1}s_i-M_1)
\delta(\sum_{i\in\Omega_2}s_i-M_2) \;\mathcal{O}
\nonumber\\* \label{defmicro}
\end{eqnarray}
we can rewrite the partition function (\ref{Zpart1}) in the form:
\begin{eqnarray}
\Z&=&\sum_{m_1,m_2}g(m_1,m_2) \nonumber\\*
& &\times \left<\exp{\left(i\frac{\theta}{2} \sum_i
  s_i + F\sum_{<ij>}{s_is_j}\right)}\right>_{m_1,m_2}. 
  \label{sumOverM}
\end{eqnarray}
The $\theta$ term in Eq. (\ref{sumOverM}) is just
$i\theta\left(m_1+m_2\right)N/4$, and therefore constant under
fixed $m_1$ and $m_2$; we can take it out of the expected value,
arriving at
\begin{eqnarray}
\Z&=&\sum_{m_1,m_2}g(m_1,m_2)e^{\frac{1}{4}Ni\theta(m_1+m_2)} \nonumber\\*
& &\times \left<\exp{\left(F\sum_{<ij>}{s_is_j}\right)}\right>_{m_1,m_2}.
\label{sumOver{m_2}}
\end{eqnarray}
We cannot evaluate exactly the expectation value in Eq.
(\ref{sumOver{m_2}}), as that would be equivalent to solving exactly
the model for arbitrary values of the external field. Instead we
perform a cumulant expansion and truncate at a given order. Let us
recall the definition:
\begin{equation}\label{CumDef1}
\left<e^{tX}\right> \equiv
\exp{\left(\sum_{n=1}^\infty\kappa_n\dfrac{t^n}{n!}\right)},
\end{equation}
where the $n$th cumulant $\kappa_n$ is an $n$th degree polynomial in
the first $n$ noncentral moments of $X$, given by the following
recursion formula:
\begin{equation}\label{cumulantDef}
\kappa_n = \mu'_n - \sum_{m=1}^{n-1}\binom{n-1}{m-1}\kappa_m\mu'_{n-m},
\qquad \mu'_n\equiv\left<X^n\right>.
\end{equation}
By expanding in cumulants in our partition function, taking $t=F$ and
$X=\sum s_is_j$, we obtain
\begin{eqnarray}
\Z&=&\sum_{m_1,m_2}g(m_1,m_2) \nonumber\\*
& &\times \exp{\left(\frac{1}{4}Ni\theta(m_1+m_2) +
  \sum_{n=1}^\infty\kappa_n(m_1,m_2)\dfrac{F^n}{n!}\right)}, \nonumber\\*
   \label{ZenKappas}
\end{eqnarray}
where now the moments are given by
\begin{equation}
\mu'_n = \left<\left(\sum_{<i,j>}s_is_j\right)^n\right>_{m_1,m_2}.
\end{equation}
The computation of these quantities is somewhat involved, and we
relegate the details to Appendix \ref{appendixcumulants}. We calculate the
cumulants using a numerical (but exact) method, up to $n=8$. The
results, at leading order in $N$\footnote{We can calculate the
  subleading terms also, but they become irrelevant as we approach the
  thermodynamic limit.}, for $d=2$, are
\begin{eqnarray}
\kappa_1 &=& 2Nm_1m_2, \nonumber \\[7pt]
\kappa_2 &=& 2N({m_1}^2 - 1)({m_2}^2 - 1), \nonumber\\[7pt]
\kappa_3 &=& 8Nm_1m_2({m_1}^2 - 1)({m_2}^2 - 1), \nonumber\\[7pt]
\kappa_4 &=& 4N(21{m_1}^2 {m_2}^2 - 9({m_1}^2 + {m_2}^2) + 5) \nonumber\\*
         & &\times ({m_1}^2 - 1)({m_2}^2 - 1), \nonumber\\[7pt]
\kappa_5 &=& 32N(51{m_1}^2{m_2}^2 - 39{m_1}^2 - 39{m_2}^2 + 31) \nonumber\\*
         & &\times m_1m_2({m_1}^2 - 1)({m_2}^2 - 1), \nonumber\\[7pt]
\kappa_6 &=& 64N(675{m_1}^4{m_2}^4 - 690[{m_1}^4{m_2}^2 + {m_1}^2{m_2}^4]
			\nonumber\\*
         & &+ 705{m_1}^2{m_2}^2 + 75[{m_1}^4 + {m_2}^4 -{m_1}^2 -{m_2}^2] + 8)
         	\nonumber \\*
         & &\times({m_1}^2 - 1)({m_2}^2 - 1), \nonumber\\[7pt]
\kappa_7 &=& 128N(10935m_1^4m_2^4 -13950[m_1^4m_2^2 + m_1^2m_2^4] \nonumber\\*
		 & &+ 3375[m_1^4 + m_2^4] + 17760m_1^2m_2^2 - 4290[m_1^2 + m_2^2]
		 \nonumber\\
		 & &+ 1051) m_1m_2(m_1^2 - 1)(m_2^2 - 1). \nonumber\\[7pt]
\kappa_8 &=& 32N(1685565{m_1}^6{m_2}^6 - 2604735[{m_1}^6{m_2}^4 \nonumber\\*
         & &+ {m_1}^4{m_2}^6]  + 994455[{m_1}^6{m_2}^2 + {m_1}^2{m_2}^6]
         \nonumber\\*
         & &- 55125[{m_1}^6 + {m_2}^6] + 4026645{m_1}^4{m_2}^4 \nonumber\\*
         & &- 1541085[{m_1}^4{m_2}^2 + {m_1}^2{m_2}^4] + 85575[{m_1}^4
            + {m_2}^4] \nonumber\\*
         & &+ 595077{m_1}^2{m_2}^2 - 33663[{m_1}^2
            + {m_2}^2] + 2125)\nonumber\\*
         & &\times(m_1^2 - 1)(m_2^2 - 1)
\label{Cumulants}
\end{eqnarray}

Now we can compute an approximation to the expected value of any
observable of the form $\mathcal{O}(m_1,m_2)$ as follows:
\begin{eqnarray}
\left< \mathcal{O} \right> &=& \dfrac{1}{\Z}
\sum_{m_1,m_2}\mathcal{O}(m_1,m_2)g(m_1,m_2) \nonumber\\*
& &\times\exp \left\lbrace
i\theta\dfrac{M_1+M_2}{2}
+\sum_{n=1}^{n_{max}}\dfrac{F^n}{n!}\kappa_n(m_1,m_2) \right\rbrace,
\nonumber\\* \label{sumExponentials}
\end{eqnarray}
where $\left< \mathcal{O} \right>$ depends implicitly on the number of
cumulants included in the approximation, $n_{max}$, and on the number
of spins of the system $N$. Taking the limit of both $n_{max}$ and $N$
to infinity, we should recover the exact result in the thermodynamic
limit. Using this technique, we have computed several observables,
such as the density of free energy $\phi$, the density of internal
energy $e$, the specific heat $c_v$ and both the usual and the
staggered magnetization $\langle m\rangle$ and $\langle m_s\rangle$,
respectively. The precise definitions of the computed observables are
the following:
\begin{eqnarray}
\phi &\equiv& -\dfrac{1}{NF}\log\Z, \\[7pt]
e &\equiv& -\dfrac{1}{2N}\dfrac{d\log\Z}{dF},
\qquad c_v\equiv -F^2\dfrac{d}{dF}e, \\[7pt]
\langle m\rangle &\equiv& \left< \dfrac{m_1 + m_2}{2}\right>,
\qquad \langle m_s\rangle \equiv \left< \dfrac{m_1 - m_2}{2} \right>.
\end{eqnarray}
It must be noted that at $\theta=\pi$, where the model has an
analytical solution, the free energy has a singularity at $F=0$
\cite{Lee:1952ig,McCoy:1967zz}. In the next section we will talk about its
nonsingular part, which is simply the result of subtracting the
singular term from the full expression:
\begin{equation}
\phi \equiv \phi_{ns} - \dfrac{1}{2F} \log{(1-e^{4F})}.
\end{equation}

As we have mentioned before, the complex-valued exponentials in Eq.
(\ref{sumExponentials}) give rise to a severe sign problem. To deal
with it we use a multiprecision algorithm, which allows us to keep as
many digits as needed. In order to crosscheck our calculations we have
used several multiprecision libraries (GMP, GNU MPFR, GNU MPC, gmpy2)
to do the sum over $m_1$ and $m_2$. The computational cost when
computing the observables grows on one hand with $N^2$ due to the
number of summands in (\ref{sumExponentials}). In addition to that,
the number of digits needed grows linearly with $N$, increasing the
cost of each multiprecision operation.

\section{Results}
\label{results}

\begin{figure}
  \includegraphics[scale=1]{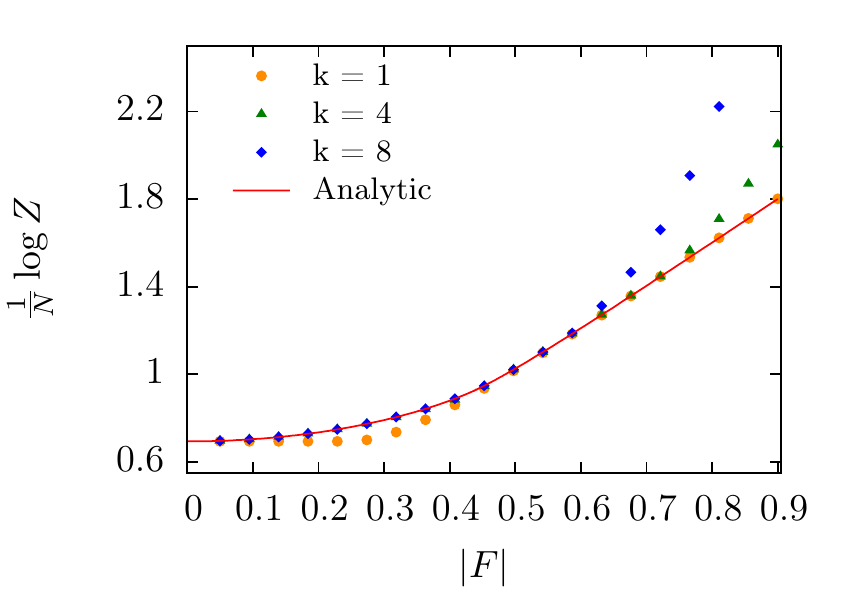}
  \caption{Free energy ($-F\phi$) at $\theta = 0, N=2000$ for the square-lattice
  AF Ising model in the $k$th cumulant approximation.}
  \label{free0}
\end{figure}

\begin{figure}
  \includegraphics[scale=1]{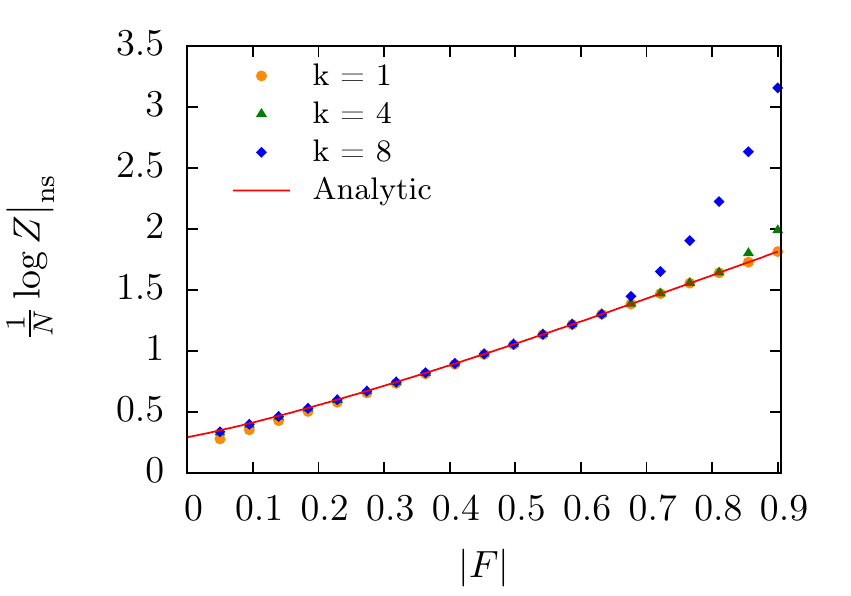}
  \caption{Nonsingular part of the free energy ($-F\phi$) at $\theta =
  \pi, N=2000$.}\label{freePI}
\end{figure}

At $\theta=0$ and $\pi$ we know the analytical solution for the
two-dimensional Ising model \cite{Onsager:1943jn, Lee:1952ig,
McCoy:1967zz}, and therefore we can compare the exact results with the
approximations obtained from Eq. (\ref{sumExponentials}). We can see in
Figs.
\ref{free0} and \ref{freePI} the density of free energy as a function
of the coupling $|F|$, for different approximations. Concretely we
show the approximations obtained by keeping only the first, up to the
fourth, and up to the eighth cumulant. For clarity we show only the results
corresponding to the largest size $N$ that we have calculated,
although we have carefully checked that the finite-size effects are
tiny at that value of $N$. We can see that the agreement with the
exact result, especially for the fourth and eighth approximations, is
excellent at small $|F|$, where we can expect the cumulant expansion
to be well behaved. At $|F| \gtrsim 0.57$ the approximations start to
drift away from the analytic result, especially the eighth, possibly
indicating the lack of convergence of the cumulant expansion at such
larger couplings.

\begin{figure}
  \includegraphics[scale=1]{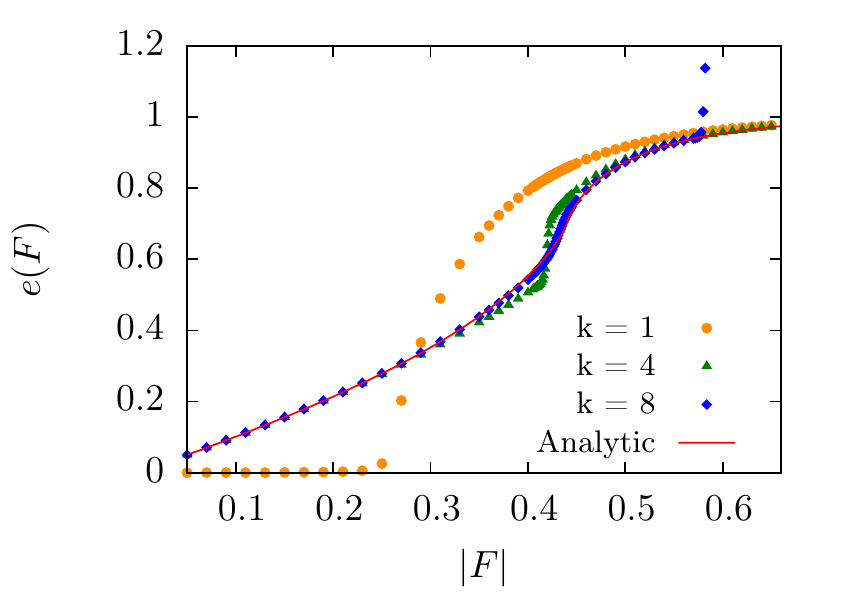}
  \caption{Internal energy $e(F)$ computed for one,
  four, and eight cumulants at $\theta=0$ and $N=2000$.}\label{eT0}
\end{figure}

\begin{figure}
  \includegraphics[scale=1]{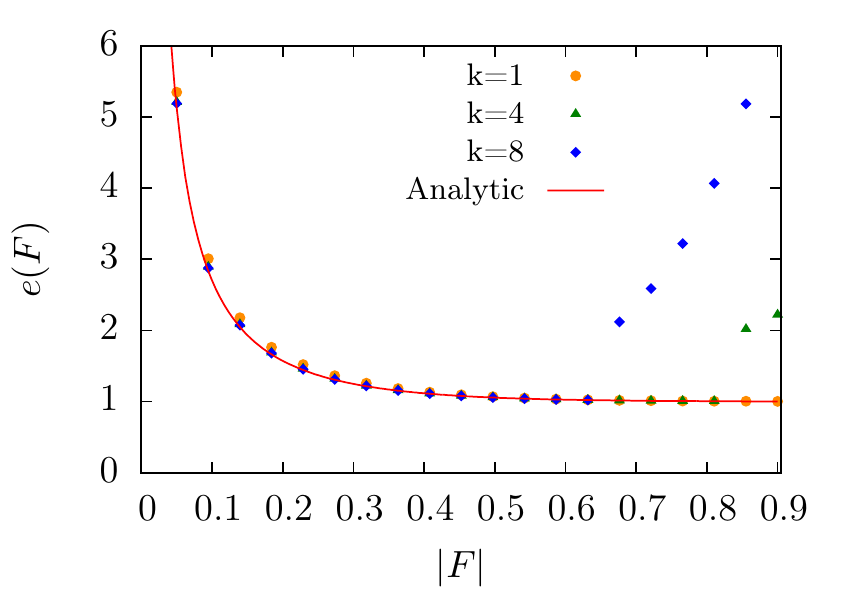}
  \caption{Internal energy density $e(F)$ at $\theta = \pi, N=2000$ at several
  cumulant expansions.}\label{energy}
\end{figure}

\begin{figure}
  \includegraphics[scale=1]{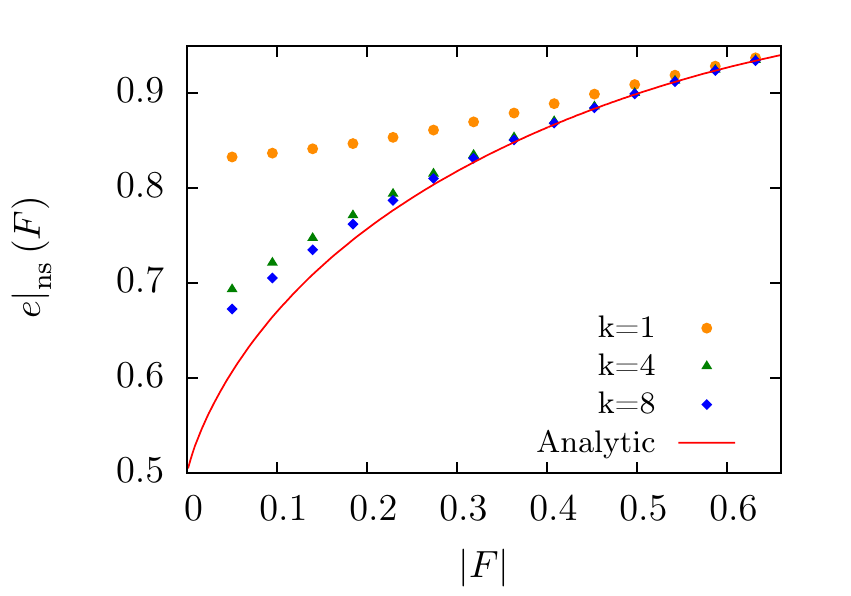}
  \caption{nonsingular part of the internal energy at $\theta = \pi, N=2000$.}
  \label{energywout}
\end{figure}

\begin{figure}
  \includegraphics[scale=1]{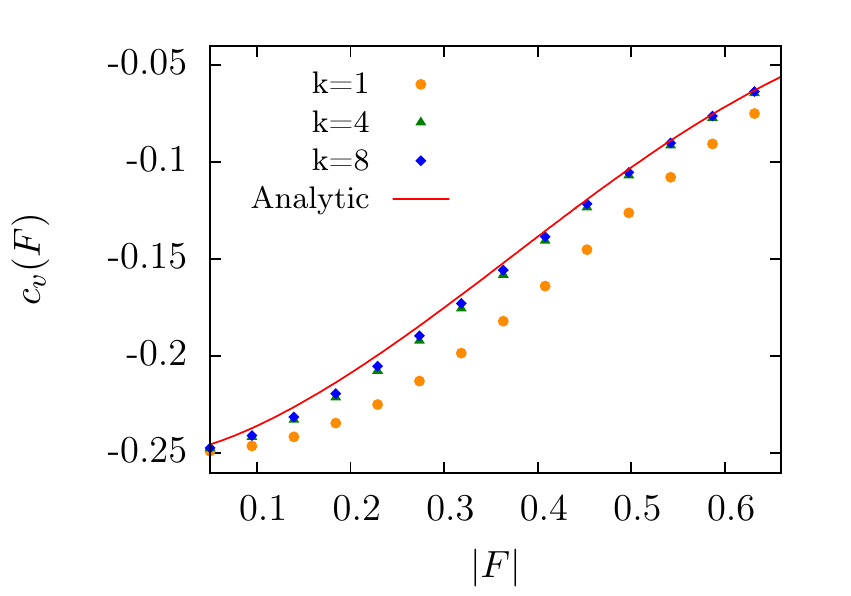}
  \caption{Specific heat at $\theta=\pi, N=2000$, plotted against the analytical
  expression.}\label{cv_pi}
\end{figure}

\begin{figure}
  \includegraphics[scale=1]{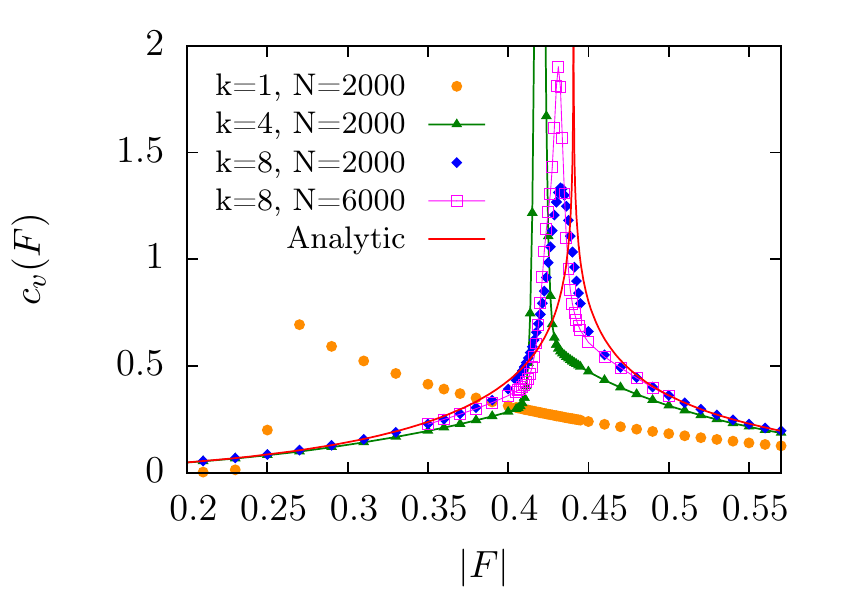}
  \caption{Specific heat at $\theta=0$, plotted against the analytical
  solution. At $\theta=0$, $F_c=\log(1+\sqrt{2})/2\approx0.4407$.}\label{cv0}
\end{figure}

The above results are consistent with those of the density of internal
energy, which we can see in Figs. \ref{eT0}-\ref{energywout}. The same can be said about the specific heat for
$\theta=\pi$, in Fig. \ref{cv_pi}. The results of the specific heat
for $\theta=0$, in Fig. \ref{cv0}, show also a good agreement with
the analytical solution, as long as we are far from the critical
point. In the neighborhood of the critical point we can see that
keeping a finite number of cumulants has a strong impact. However, the
results seem to converge to the exact solution quickly when we
increase the number of cumulants, and indeed the peak when including
all eight cumulants is not far from the analytic result.

The agreement with the exact results both at $\theta = 0$ and at
$\pi$ suggests that the cumulant expansion can be trusted at
all values of $\theta$, as long as $|F| \lesssim0.57$.

We expect a nonvanishing value of $\langle m_s\rangle$ to signal the
transition from the paramagnetic to the staggered phase. Because of
translational symmetry, we cannot simply compute this observable,
since for a finite $N$ system it is always zero [permuting $m_1$ and
$m_2$ leaves Eq. (\ref{sumExponentials}) invariant]. However, we can
compute $\langle m_s^2\rangle$, which also separates the weak and
strong coupling phases.

\begin{figure}    
  \includegraphics[scale=1]{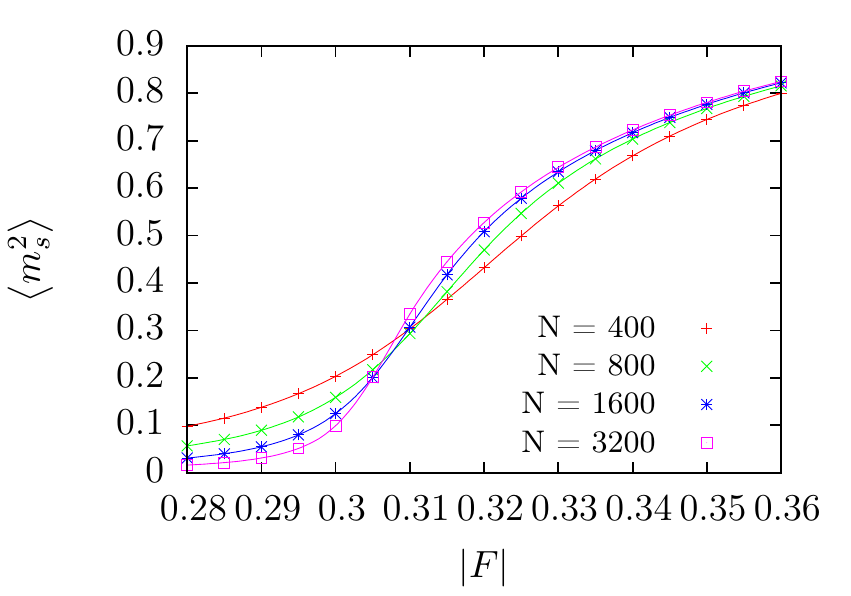}
  \caption{$\langle m_s^2\rangle$ curves at $\theta=2,k=8$. Solid lines are
  just a guide to the eye.}\label{msT2}
\end{figure}

\begin{figure}    
  \includegraphics[scale=1]{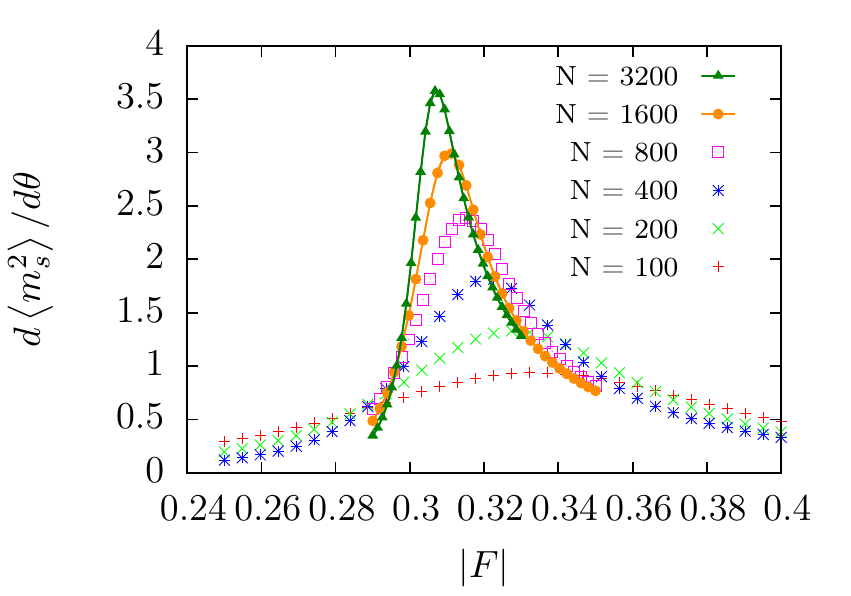}
  \caption{Scaling of $d\langle m_s^2\rangle/d\theta$ at
  $\theta=2,k=8$. Solid lines are a guide to the eye.}\label{scaling}
\end{figure}

In Fig. \ref{msT2} we show results for $\langle m_s^2\rangle$ at
$\theta=2$.  One can see how, as we approach the thermodynamic limit,
$\langle m_s^2\rangle$ becomes a steeper function of $|F|$. To obtain
the critical line for a given cumulant approximation, we numerically
calculate the quantity $\frac{d}{d\theta} \langle m_s^2\rangle$ (which
should diverge in the thermodynamic limit at the critical line), and
find the maximum along lines of constant $\theta$. This gives us, for
each size $N$ and each value of $\theta$, $F_c(\theta)$. We can see in
Fig. \ref{scaling} the behavior of such quantity as a function of
$F$ and $N$, for the specific value $\theta = 2$, in the eight
cumulant approximation. The height of the peak does not scale as $N$,
at least at the volumes we have been able to calculate, therefore
suggesting a continuous phase transition; however, our data are not
extensive enough to calculate the critical exponents.

\begin{figure}    
  \includegraphics[scale=1]{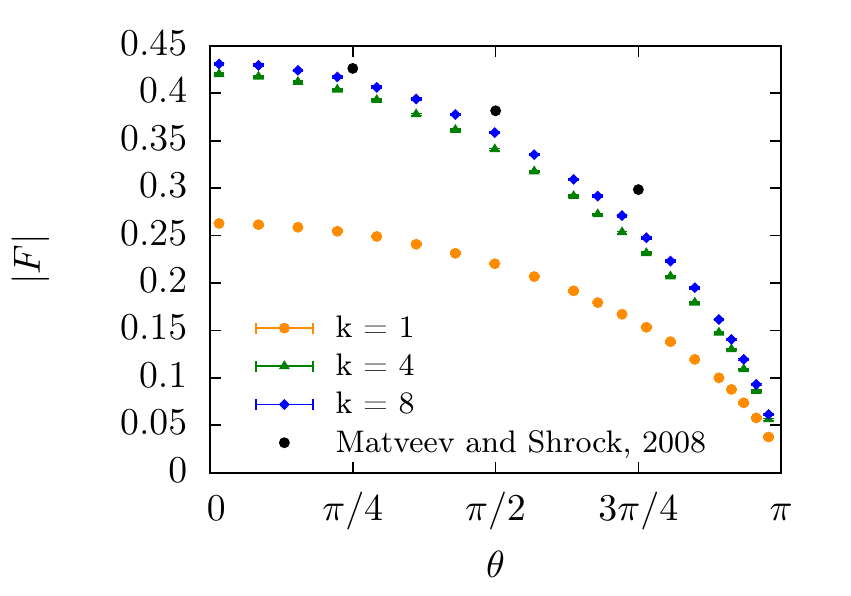}
  \caption{The critical line $F_c(\theta)$, computed as the maximum of
    $d\langle m_s^2\rangle/d\theta$ at $N=2000$. The maximal
    $F$ points obtained in \cite{Shrock:2008} are also
    shown.}\label{trline}
\end{figure}

The phase diagram obtained in this way is shown in Fig.
\ref{trline}, for several truncation orders of the cumulant expansion.
The transition lines that we obtain lie entirely below $|F| = .45$,
where we have good evidence that the cumulant expansion works
well. The change from the line corresponding to $k=1$ and $4$ is
very large, but the results seem to stabilize quickly with the order
of the expansion, and the lines corresponding to $k=4$ and $8$ are
quite close together. Therefore we expect the phase diagram for $k=8$
to be a quite accurate approximation to the exact one. Further
evidence of this is the agreement with the few maximal values for
$F_c$ estimated in \cite{Shrock:2008} from the computation of the
zeros of the partition function of the model in the complex
temperature-magnetic field plane. As can be seen in the plot, they lie
above but quite close to our $k=8$ line.

\begin{figure}
  \includegraphics[scale=1]{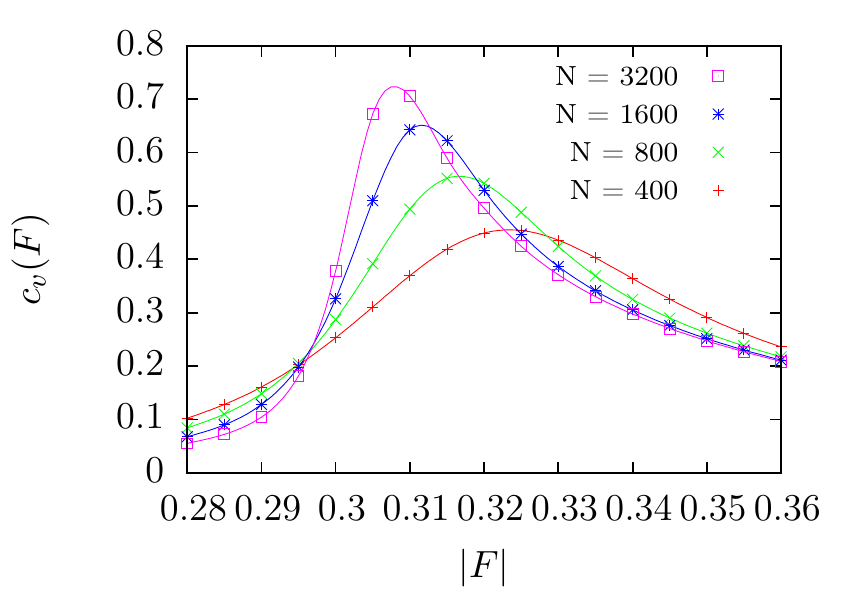}
  \caption{Specific heat $c_v$ with $k=8$ and $\theta=2$. Solid lines
  are just a guide to the eye.}\label{cvT2}
\end{figure}

As another crosscheck we show in Fig. \ref{cvT2} results for the
specific heat at $\theta=2$ in the eight cumulant approximation,
computed for several system sizes. The behavior is similar to the one
in Fig. \ref{scaling}: a peak of increasing height in the vicinity
of the critical point, and smooth behavior and small finite N effects
elsewhere.

\section{Conclusions}
\label{conclusions}

We have analyzed the two-dimensional antiferromagnetic Ising model
with an imaginary magnetic field by analytical techniques. We have
calculated the first eight cumulants of what is essentially the
expansion of the effective Hamiltonian in powers of the inverse
temperature, and computed physical quantities for a large number of
degrees of freedom with the help of multiprecision algorithms. The
motivation for such a calculation was to have an example of a physical
system with SSP and nontrivial phase structure, the dynamics of which
is well known, at least in the high-temperature region.

Our results confirm the qualitative picture described in
\cite{Azcoiti:2011ei}, and predict the existence of two phases in this
model, which can be characterized by the staggered magnetization as an
order parameter. The finite-size scaling suggests that the two phases
are separated by a continuous phase transition line. The position of
the critical point at $\theta = 0$ is in very good agreement with the
exact result $F_c=\log(1+\sqrt{2})/2\approx0.4407$, and the free and internal energy densities at $\theta =
\pi$ agree also well with the analytical prediction, at least in the
high-temperature regime, thus giving reliability to our results in
this region. Therefore this model could be a good laboratory to check
proposals to simulate physical systems afflicted by a SSP.

\begin{acknowledgments}
This work was funded by Ministerio de Economía y Competitividad/Fondo
Europeo de Desarrollo Regional Grants No. FPA2012-35453 and No.
FPA2015-65745-P and Diputación General de Aragón-Fondo Social Europeo
Grant No. 2015-E24/2.
\end{acknowledgments}

\appendix
\section{Computation of the cumulants $\kappa_n$}
\label{appendixcumulants}
In order to use expressions (\ref{ZenKappas}) and
(\ref{sumExponentials}), we need to compute the cumulants
$\kappa_n$. The $n$th cumulant can be calculated in terms of the
first $n$ noncentral moments $\mu'_n$,
\begin{equation}\label{momentDef}
\mu'_n\equiv \left<\left(\sum_{<ij>}s_is_j\right)^n\right>_{m_1,m_2},
\end{equation}
by means of the recursion relation (\ref{cumulantDef}). The summation
over $<ij>$ runs over each couple of neighboring spins, or in other
words, over each link. Two neighboring spins always belong to
different sublattices.

Before going further, let us comment on two intermediate
results. First, we consider a lattice of $N$ spins, the
magnetization of which is the sum $m=\sum_i s_i$,
and ask about the expected
value of the product of $n$ of these spins at fixed $m$ (or fixed
$N_+$, the number of positive spins), that is, $\langle s_1s_2\cdots
s_n\rangle_{m}$. One can perform this calculation by means of the
microcanonical formalism, arriving at
\begin{equation}\label{s1sn}
\langle s_1s_2\cdots
s_n\rangle_{m}=\dfrac{1}{\binom{N}{N_+}}\sum_{k=0}^n(-1)^k
\binom{n}{k}\binom{N-n}{N_+-n+k}.
\end{equation}
In the above expression, $k$ can be read as the number of negative
spins in the product $s_1s_2\cdots s_n$. In this way, the first
summand, $k=0$, counts the number of states with zero negative spins
in the product $s_1s_2\cdots s_n$
and multiplies it by the expected value of the product in this case,
$(-1)^0=1$. The second one, $k=1$, does the same for one negative spin
in $s_1\cdots s_n$, and so on. Dividing the sum by the total number of
configurations with magnetization $m=2N_+/N-1$, one obtains the
previous expected value at fixed $m$. Secondly, consider an observable
$\mathcal{O}(m_1,m_2)$ in our two sublattice system, with a dependence
on $m_1$ and $m_2$ such as we can write it as
$\mathcal{O}_1(m_1)\mathcal{O}_2(m_2)$. In this case, from the
definition (\ref{defmicro}) of the expected value at fixed $m_1$ and
$m_2$, we have
\begin{equation}
\left<\mathcal{O}_1(m_1)\mathcal{O}_2(m_2)\right>_{m_1,m_2} =
\left<\mathcal{O}_1(m_1)\right>_{m_1}\left<\mathcal{O}_2(m_2)\right>_{m_2}.
\end{equation}
This immediately applies to the spin product $s_1s_2\cdots s_n$. We
can always divide it into two products $s_a\cdots s_b$ and
$s_\alpha\cdots s_\beta$, each one containing the spins of one of
the sublattices, and then
\begin{equation}\label{separable}
\left<s_1s_2\cdots s_n\right>_{m_1,m_2}=\left<s_a\cdots
s_b\right>_{m_1} \left<s_\alpha\cdots s_\beta\right>_{m_2}.
\end{equation}
With the previous couple of results, we come back to Eq.
(\ref{momentDef}), and apply the linearity of the expected value,
arriving at
\begin{equation}\label{geomFact}
\mu'_n=\sum_{<ij>,<kl>,\cdots,<pq>}\left<s_is_js_ks_l\cdots
s_ps_q\right>_{m_1,m_2},
\end{equation}
which is the sum of the expected values of the product of $n$ links,
running over all permutations with repetitions of these links. Then,
in every summand we have the product of $2n$ spins, in some cases with
some of them identical. Taking into account that $s_i^2=1\:\forall i$,
each summand can be reduced to the expected value of the product of
$n_1+n_2$ different spins, $n_1$ and $n_2$ being the number of spins
in each sublattice. Since by means of Eq. (\ref{s1sn}) we already have an
expression that computes $\langle s_1\cdots s_n\rangle_{m}$, the problem
is reduced to count how many summands in Eq. (\ref{geomFact}) have $(n_1,n_2)$
spins. We call these numbers \textit{geometrical factors}, and denote
them by $\mathcal{G}(n_1,n_2)$. Following this convention, we can
write the $n$th central moment as
\begin{equation}\label{sumaSymb}
\mu'_n=\sum_{\{n_1,n_2\}}\mathcal{G}(n_1,n_2) \langle\underbrace{s_a\cdots
  s_b}_{n_1\:\text{spins}}\rangle_{m_1} \langle\underbrace{s_\alpha\cdots
  s_\beta}_{n_2\:\text{spins}}\rangle_{m_2},
\end{equation}
where the sum runs over the couples of integers $(n_1,n_2)$ the sum of
which is even and less than or equal to $n$.

The computation of the geometrical factors $\mathcal{G}(n_1,n_2)$ can
be done by hand for the first few cumulants. As an example, for the
second noncentral moment $\mu'_2$ we have to compute four cases: the
two links being the same (sharing both spins), sharing only one spin
belonging to the first or the second sublattice, and finally not
sharing any spin at all. That is, in terms of the previous notation,
\begin{equation}
\{(n_1,n_2)\} = \{(0,0),(2,0),(0,2),(2,2)\}.
\end{equation}
The factors $\mathcal{G}(n_1,n_2)$ can be computed easily in this
case, even for an hypercubic lattice of arbitrary dimension $d$,
arriving at the following expression for the second moment
\begin{eqnarray}
\mu'_2 &=& Nd\langle 1\rangle + Nd(d-1)(\langle s_1s_2\rangle_{m_1}
+\langle s_1s_2\rangle_{m_2}) \nonumber\\*[7pt]
& &+ Nd(Nd-2(d-1)-1)\langle s_1s_2\rangle_{m_1}\langle s_1s_2\rangle_{m_2}.
\nonumber\\*
\end{eqnarray}
We can use this expression to calculate the second cumulant
$\kappa_2$,
\begin{equation}
\kappa_2=\mu'_2-{\mu'}_1^2\xrightarrow{N\to\infty} Nd(m_1^2-1)(m_2^2-1),
\end{equation}
where we have taken the thermodynamic limit, keeping only the terms of
order $\mathcal{O}(N)$, which is the leading order for all
cumulants. Subleading orders can be preserved if needed, but they are
not relevant for our paper. The difficulty of the previous computation
escalates quickly with the order $n$ of the cumulant, and it is quite
cumbersome for just $n\geq4$. In order to get beyond this limitation,
we have developed a program which computes the geometrical factors
$\mathcal{G}(n_1,n_2)$ numerically for a finite $L\times L$
bidimensional lattice. Since these factors $\mathcal{G}(n_1,n_2)$ are
polynomials in $N$ of order $\leq n$ (and with integer coefficients),
we can run the program for lattices of $n+1$ different sizes,
obtaining a set of ($N$,$\mathcal{G}(N)$) points, which we can use to
recover the exact integer coefficients of each geometrical factor, by
means of the Lagrange interpolation formula.

The basic idea of the program is very simple. We just construct a
periodic rectangular $L\times M$ lattice, with $L,M>n$, $n$ being the
order of the cumulant we want to compute. With this restriction we
avoid products of links crossing the entire lattice, that would not
appear in the thermodynamic limit for any finite cumulant. Once we
have this, we start a loop running over all the permutations with
repetitions of $n$ links, and perform the following steps,
\begin{itemize}
\item We have a product of $n$ links, or equivalently $2n$ spins,
$s_1\cdots s_{2n}$.
\item Recursively, we remove couples of equal spins from this product.
\item We classify the remaining product by the number of spins in each
sublattice, $(n_1,n_2)$.
\item We add one to the geometric factor $\mathcal{G}(n_1,n_2)$ and proceed
to the next iteration.
\end{itemize}
When the algorithm finishes, we obtain all the $\mathcal{G}(n_1,n_2)$
values for a given $N=LM$. The computational cost is associated to the
number of iterations of the main loop, which grows as ${(LM)}^{n}$,
that is, exponentially with the order of the cumulant. In practice, we
have only reached the computation of the fourth cumulant with this
program. However, a number of optimizations can be implemented in
order to reach higher order cumulants, which we summarize in what
follows.
\subsection{Translational symmetry}
  Our lattice is symmetric under
  translations, implying that all geometrical factors are proportional
  to $Nd$, the number of links. Fixing, e.g., the first link of the
  product, one obtains the same $\mathcal{G}(n_1,n_2)$, but divided by
  a common factor $Nd$. The same factor is gained in the overall speed
  of the program. In addition to that, the degree of the polynomials
  $\mathcal{G}(n_1,n_2)$ is also reduced by one, and it suffices with
  $n$ (instead of $n+1$) different sizes in order to recover the $N$
  dependence. One can go even further by realizing that the
  geometrical factor corresponding to non-neighboring links,
  $\mathcal{G}(n,n)$, is the only one with maximum degree
  $N^{n-1}$. This allows us to express it in terms of the remaining
  factors,
\begin{eqnarray}
\dfrac{1}{Nd}\mathcal{G}(n,n) &=& (Nd)^{n-1} \nonumber\\*
& &-\dfrac{1}{Nd}\sum_{\{(n_1,n_2)\}\backslash(n,n)}\mathcal{G}(n_1,n_2),
\end{eqnarray}
which are only of order $n-2$ or less. This means that it is enough to
run the program for $n-1$ lattice sizes, compute all the geometrical
factors but $\mathcal{G}(n,n)$ via the Lagrange interpolator, and then
with the previous expression find the $N$ dependence of this last
factor.

\subsection{From permutations to combinations}
  The product of links
  commutes, so its contribution to the geometrical factors is the same
  regardless of the order. Then, we can change the main loop over
  permutations with repetition to a loop over combinations with
  repetition, by taking into account the multiplicity of each
  combination. Schematically, we perform
\begin{eqnarray}
&\sum_{i,j,\dots,k}&\text{contrib}(l_il_j\cdots l_k) \nonumber\\*
& &\rightarrow \sum_{i\leq j\leq\cdots\leq k}\text{mult}
\times\text{contrib}(l_il_j\cdots l_k),
\end{eqnarray}
where \textit{contrib} represents a function in our program that takes
a product of links and returns the contribution to the geometrical
factors. If there are $r$ different links, each one appearing
$k_1,\dots,k_r$ times, the multiplicity of the combination is given by
\begin{equation}
\text{mult}=\dfrac{n!}{k_1!\cdots k_r!}.
\end{equation}

\subsection{Blocks - Grouping links together}
  Many of the link
  products have few, if any, repeated spins, and their contributions
  to the geometrical factors can be counted without having to analyze
  one by one each of them. This is possible by grouping them in sets of
  links that we will call in what follows \textit{blocks}, and
  replacing the loop over link products by a loop over block
  products. When the blocks in a product are not neighbors (i.e., they
  do not have any common spin), we do not need to perform the
  computation link by link and the contribution can be summed up
  trivially. Let $b_1$ and $b_3$ be two non-neighboring blocks, each
  one composed by $N_b$ links, and let us denote the contributions to
  the geometrical factors by $\lambda(n_1,n_2)$, where $\lambda$ is an
  integer counting how many products of links have $n_1$ ($n_2$) spins
  in the first (second) sublattice. Then we have
\begin{equation}
contrib(b_1b_3)=N_b^2(2,2),
\end{equation}
or in general, for the product of $k$ non-neighboring blocks,
$N_b^k(k,k)$. Following this strategy, we divide our lattice into
\textit{unidimensional} blocks of $2M$ links, in a way that the $j$th
block, $b_j$, contains all links the first spin of which belongs to the
$j$th column. As a consequence, $b_j$ is a neighbor of blocks $j-1$ and
$j+1$, and, taking into account the boundary conditions, $b_0$ and
$b_{L-1}$ are neighbors too.

When we have a product of neighboring blocks, we proceed as before,
analyzing the link products one by one, and there is no computational
saving. But when the $n$ blocks are not neighbors, we move from
$(Nd)^n$ iterations to a single one.

\subsection{Clusters of blocks} The block method, as defined
  above, fails to save any computation time if two or more blocks are
  neighbors in a given block product. However, we can extend the
  method by dividing each block product into several subproducts, which
  we will denote as \textit{clusters}. In each cluster, one can always
  connect one block to another by the equivalence relation of being
  neighbors (sharing spins). And in the same way, in each product
  different clusters never share any spin. This allows us to compute
  the contributions of each cluster separately, and then compose them
  with the following law,
\begin{equation}
\lambda_{1}(a,b)\oplus\lambda_{2}(c,d)=\lambda_1\lambda_2(a+c,b+d).
\end{equation}
If the contributions of the clusters involve more than one geometrical
factor, linearity applies,
\begin{eqnarray}
\sum_{ab}\lambda_{ab}(a,b) &\oplus& \sum_{cd}\lambda_{cd}(c,d)=
\nonumber\\* & &\sum_{ab,cd}\lambda_{ab}\lambda_{cd}(a+c,b+d).
\label{compositionLaw}
\end{eqnarray}
Processing one cluster with $k$ blocks takes a computing time
proportional to $(Nd)^k$. So dividing the whole block product in
smaller clusters implies for almost every block product a significant
amount of time saved. Only when all the blocks are part of the same
cluster there is no speed up.

Another major optimization can be performed by realizing that
translational invariance can also be applied here, since a given
cluster, say $b_0b_1b_1$, and any of its translations,
$b_{0+t}b_{1+t}b_{1+t}$, have the same contribution to the geometrical
factors. Then, when a cluster is going to be computed, we can express
it in terms of its equivalence class, compute its contribution, and
store it in memory. Every time one of its translations appears, we
just take the value from the memory, saving a lot of computing
time. In addition to that, once we have computed the factors
$\mathcal{G}(n_1,n_2)$ for the first size $L\times M$, we know in
advance \textit{all} the cluster contributions for any $L'\times M$
lattice (the blocks keep its size constant). Since almost all the
computing time is spent in figuring out the cluster contributions, we
reduce in this way the full problem of computing the geometrical
factors in lattices of $n-1$ different sizes to \textit{only} one
size, the smallest one, $M\times M$. In practice, the time spent by
the rest of the sizes needed is barely the $1-2\%$ of that of the
first size.

\subsection{Computation of a cluster}
  The last optimization
  concerns the computation of the clusters themselves. Until now it is
  done simply by performing a loop over each possible permutation of
  links belonging to each of the blocks in the cluster. However, one
  can go one step further and divide the blocks composing the cluster
  into smaller sets, that we will call \textit{sites}. A site is simply
  the set of two links the first spin of which lies in the site $i,j$,
  that is,
\begin{equation}
\text{site}(i,j)\equiv\{s_{ij}s_{i+1,j},s_{ij}s_{i,j+1}\}.
\end{equation}
With this new subdivision, we can apply in the same way the techniques
described above. In order to compute the cluster $b_1\dots b_k$, we
start a loop over every permutation of sites $s_1\dots s_k$, with
$s_i\in b_i$. Each site product is divided into clusters, the
contributions of which can be summed with Eq. (\ref{compositionLaw})
and are
calculated by performing another loop over each link product ($2^k$
iterations for a site product of $k$ elements). Finally, by summing up
each site product contribution, we obtain the whole cluster
contribution.

All the described optimizations do not remove the exponential
dependence on $n$ of the algorithm. However, they allow us to reach
the eighth cumulant, which takes about three days of computing time in a
modern laptop.

\section{Numerical tables}
\label{appendixtables}
In this appendix we present some of the data corresponding to the
figures in Sec. \ref{results}. In addition to that, we provide
numerical results for several observables at $\theta=2$ and $k=8$.

\squeezetable
\begin{table}
\caption{Numerical data for $\theta = 0$, Fig. \ref{free0}.}
\begin{ruledtabular}
  \begin{tabular}{cccc}
    $|F|$ & $-F\phi(k=1)$ & $-F\phi(k=4)$ & $-F\phi(k=8)$\\
	\hline
	0.0500 & 0.693 & 0.696 & 0.696 \\
	0.0947 & 0.693 & 0.702 & 0.702 \\
	0.1395 & 0.693 & 0.713 & 0.713 \\
	0.1842 & 0.693 & 0.728 & 0.728 \\
	0.2289 & 0.694 & 0.748 & 0.748 \\
	0.2737 & 0.700 & 0.773 & 0.773 \\
	0.3184 & 0.736 & 0.803 & 0.804 \\
	0.3632 & 0.792 & 0.840 & 0.842 \\
	0.4079 & 0.860 & 0.883 & 0.888 \\
	0.4526 & 0.935 & 0.945 & 0.947 \\
	0.4974 & 1.015 & 1.021 & 1.021 \\
	0.5421 & 1.098 & 1.101 & 1.102 \\
	0.5868 & 1.183 & 1.185 & 1.188 \\
	0.6316 & 1.270 & 1.271 & 1.312 \\
	0.6763 & 1.358 & 1.358 & 1.466 \\
	0.7211 & 1.446 & 1.446 & 1.660 \\
	0.7658 & 1.534 & 1.565 & 1.908 \\
	0.8105 & 1.623 & 1.709 & 2.224 \\
	0.8553 & 1.712 & 1.870 & 2.630 \\
	0.9000 & 1.801 & 2.049 & 3.152 \\
  \end{tabular}
\end{ruledtabular}
\end{table}

\squeezetable
\begin{table}
\caption{Numerical data for $\theta = \pi$, Fig. \ref{freePI}.}
\begin{ruledtabular}
  \begin{tabular}{cccc}
	$|F|$ & $-F\phi_{ns}(k=1)$ & $-F\phi_{ns}(k=4)$
	& $-F\phi_{ns}(k=8)$\\
	\hline
	0.0500 & 0.277 & 0.329 & 0.335 \\
	0.0947 & 0.352 & 0.392 & 0.397 \\
	0.1395 & 0.427 & 0.458 & 0.461 \\
	0.1842 & 0.503 & 0.526 & 0.528 \\
	0.2289 & 0.579 & 0.596 & 0.598 \\
	0.2737 & 0.655 & 0.668 & 0.669 \\
	0.3184 & 0.733 & 0.742 & 0.743 \\
	0.3632 & 0.811 & 0.817 & 0.818 \\
	0.4079 & 0.890 & 0.894 & 0.895 \\
	0.4526 & 0.970 & 0.973 & 0.973 \\
	0.4974 & 1.051 & 1.053 & 1.053 \\
	0.5421 & 1.133 & 1.134 & 1.134 \\
	0.5868 & 1.215 & 1.216 & 1.216 \\
	0.6316 & 1.299 & 1.299 & 1.300 \\
	0.6763 & 1.383 & 1.383 & 1.446 \\
	0.7211 & 1.468 & 1.468 & 1.650 \\
	0.7658 & 1.554 & 1.554 & 1.903 \\
	0.8105 & 1.640 & 1.640 & 2.223 \\
	0.8553 & 1.726 & 1.800 & 2.632 \\
	0.9000 & 1.813 & 1.986 & 3.155 \\
  \end{tabular}
\end{ruledtabular}
\end{table}

\squeezetable
\begin{table}
\caption{Numerical data for $\theta = \pi$, Fig. \ref{energy}.}
\begin{ruledtabular}
  \begin{tabular}{cccc}
	$|F|$ & $e(k=1)$ & $e(k=4)$ & $e(k=8)$\\
	\hline
	0.0500 & 5.350 & 5.210 & 5.189 \\
	0.0947 & 3.008 & 2.893 & 2.877 \\
	0.1395 & 2.180 & 2.086 & 2.073 \\
	0.1842 & 1.765 & 1.689 & 1.680 \\
	0.2289 & 1.521 & 1.461 & 1.455 \\
	0.2737 & 1.364 & 1.318 & 1.313 \\
	0.3184 & 1.258 & 1.223 & 1.220 \\
	0.3632 & 1.184 & 1.159 & 1.156 \\
	0.4079 & 1.132 & 1.114 & 1.112 \\
	0.4526 & 1.095 & 1.082 & 1.080 \\
	0.4974 & 1.068 & 1.059 & 1.058 \\
	0.5421 & 1.048 & 1.042 & 1.041 \\
	0.5868 & 1.034 & 1.030 & 1.030 \\
	0.6316 & 1.024 & 1.022 & 1.021 \\
	0.6763 & 1.017 & 1.016 & 2.121 \\
	0.7211 & 1.012 & 1.011 & 2.588 \\
	0.7658 & 1.009 & 1.008 & 3.222 \\
	0.8105 & 1.006 & 1.006 & 4.068 \\
	0.8553 & 1.004 & 2.019 & 5.185 \\
	0.9000 & 1.003 & 2.220 & 6.638 \\
  \end{tabular}
\end{ruledtabular}
\end{table}

\squeezetable
\begin{table}
\caption{Numerical data for $\theta = \pi$, Fig. \ref{energywout}.}
\begin{ruledtabular}
  \begin{tabular}{cccc}
	$|F|$ & $e|_{ns}(k=1)$ & $e|_{ns}(k=4)$ & $e|_{ns}(k=8)$\\
	\hline
	0.0500 & 0.833 & 0.693 & 0.672 \\
	0.0947 & 0.837 & 0.721 & 0.705 \\
	0.1395 & 0.841 & 0.747 & 0.735 \\
	0.1842 & 0.847 & 0.771 & 0.762 \\
	0.2289 & 0.854 & 0.794 & 0.787 \\
	0.2737 & 0.861 & 0.815 & 0.810 \\
	0.3184 & 0.870 & 0.835 & 0.831 \\
	0.3632 & 0.879 & 0.853 & 0.851 \\
	0.4079 & 0.889 & 0.870 & 0.869 \\
	0.4526 & 0.899 & 0.886 & 0.885 \\
	0.4974 & 0.909 & 0.900 & 0.899 \\
	0.5421 & 0.919 & 0.913 & 0.912 \\
	0.5868 & 0.929 & 0.925 & 0.924 \\
	0.6316 & 0.937 & 0.935 & 0.934 \\
	0.6763 & 0.946 & 0.944 & 2.049 \\
	0.7211 & 0.953 & 0.952 & 2.529 \\
	0.7658 & 0.960 & 0.959 & 3.173 \\
	0.8105 & 0.965 & 0.965 & 4.027 \\
	0.8553 & 0.970 & 1.985 & 5.151 \\
	0.9000 & 0.975 & 2.192 & 6.610 \\
  \end{tabular}
\end{ruledtabular}
\end{table}

\squeezetable
\begin{table}
\caption{Numerical data for the phase diagram of Fig. \ref{trline}.}
\begin{ruledtabular}
  \begin{tabular}{cccc}
	$\theta$ & $F_c (k=1)$ & $F_c (k=4)$ & $F_c (k=8)$\\
	\hline
	0.050000 & 0.263 & 0.420 & 0.431 \\
	0.266667 & 0.261 & 0.417 & 0.430 \\
	0.483333 & 0.259 & 0.412 & 0.424 \\
	0.700000 & 0.255 & 0.404 & 0.417 \\
	0.916667 & 0.249 & 0.393 & 0.406 \\
	1.133333 & 0.241 & 0.378 & 0.394 \\
	1.350000 & 0.231 & 0.361 & 0.378 \\
	1.566667 & 0.220 & 0.341 & 0.358 \\
	1.783333 & 0.207 & 0.317 & 0.335 \\
	2.000000 & 0.192 & 0.292 & 0.309 \\
	2.133333 & 0.179 & 0.272 & 0.292 \\
	2.266667 & 0.167 & 0.253 & 0.271 \\
	2.400000 & 0.153 & 0.231 & 0.248 \\
	2.533333 & 0.138 & 0.207 & 0.223 \\
	2.666667 & 0.119 & 0.179 & 0.195 \\
	2.800000 & 0.100 & 0.148 & 0.162 \\
	2.868319 & 0.088 & 0.130 & 0.141 \\
	2.936637 & 0.074 & 0.109 & 0.119 \\
	3.004956 & 0.058 & 0.086 & 0.093 \\
	3.073274 & 0.038 & 0.056 & 0.061 \\
  \end{tabular}
\end{ruledtabular}
\end{table}

\squeezetable
\begin{table}
\caption{Numerical data for $\theta=2$ and $N=400$, eight cumulants.}
\begin{ruledtabular}
  \begin{tabular}{cccccc}
	$|F|$ & $-F\phi$ & $	-i\langle m\rangle$ &
	$\langle m_s^2\rangle$ & $e$ & $c_v$\\
	\hline
	0.280 & 0.5954 & 0.3406 & 0.0976 & 0.6847 & 0.1021 \\
	0.285 & 0.6023 & 0.3315 & 0.1153 & 0.6919 & 0.1278 \\
	0.290 & 0.6093 & 0.3218 & 0.1379 & 0.7006 & 0.1607 \\
	0.295 & 0.6163 & 0.3113 & 0.1668 & 0.7111 & 0.2023 \\
	0.300 & 0.6235 & 0.2998 & 0.2034 & 0.7239 & 0.2529 \\
	0.305 & 0.6308 & 0.2872 & 0.2488 & 0.7393 & 0.3106 \\
	0.310 & 0.6383 & 0.2734 & 0.3033 & 0.7573 & 0.3693 \\
	0.315 & 0.6460 & 0.2585 & 0.3655 & 0.7775 & 0.4190 \\
	0.320 & 0.6538 & 0.2429 & 0.4323 & 0.7991 & 0.4492 \\
	0.325 & 0.6619 & 0.2273 & 0.4994 & 0.8210 & 0.4541 \\
	0.330 & 0.6703 & 0.2122 & 0.5626 & 0.8418 & 0.4359 \\
	0.335 & 0.6788 & 0.1982 & 0.6192 & 0.8608 & 0.4028 \\
	0.340 & 0.6875 & 0.1854 & 0.6681 & 0.8776 & 0.3638 \\
	0.345 & 0.6963 & 0.1738 & 0.7096 & 0.8923 & 0.3255 \\
	0.350 & 0.7053 & 0.1633 & 0.7447 & 0.9051 & 0.2910 \\
	0.355 & 0.7144 & 0.1537 & 0.7744 & 0.9162 & 0.2612 \\
	0.360 & 0.7236 & 0.1450 & 0.7999 & 0.9259 & 0.2357 \\
  \end{tabular}
\end{ruledtabular}
\end{table}

\squeezetable
\begin{table}
\caption{Numerical data for $\theta=2$ and $N=3200$, eight cumulants.}
\begin{ruledtabular}
  \begin{tabular}{cccccc}
	$|F|$ & $-F\phi$ & $	-i\langle m\rangle$ & $\langle m_s^2\rangle$
	& $e$ & $c_v$\\
	\hline
	0.280 & 0.5928 & 0.3513 & 0.0161 & 0.6633 & 0.0551 \\
	0.285 & 0.5995 & 0.3438 & 0.0214 & 0.6672 & 0.0720 \\
	0.290 & 0.6062 & 0.3360 & 0.0311 & 0.6724 & 0.1042 \\
	0.295 & 0.6129 & 0.3269 & 0.0513 & 0.6804 & 0.1815 \\
	0.300 & 0.6198 & 0.3143 & 0.0991 & 0.6954 & 0.3780 \\
	0.305 & 0.6269 & 0.2947 & 0.2013 & 0.7243 & 0.6725 \\
	0.310 & 0.6343 & 0.2705 & 0.3347 & 0.7622 & 0.7053 \\
	0.315 & 0.6421 & 0.2488 & 0.4447 & 0.7954 & 0.5897 \\
	0.320 & 0.6502 & 0.2305 & 0.5276 & 0.8222 & 0.4958 \\
	0.325 & 0.6585 & 0.2148 & 0.5920 & 0.8443 & 0.4244 \\
	0.330 & 0.6671 & 0.2010 & 0.6437 & 0.8628 & 0.3710 \\
	0.335 & 0.6758 & 0.1887 & 0.6865 & 0.8786 & 0.3302 \\
	0.340 & 0.6846 & 0.1775 & 0.7226 & 0.8923 & 0.2976 \\
	0.345 & 0.6936 & 0.1673 & 0.7537 & 0.9044 & 0.2705 \\
	0.350 & 0.7027 & 0.1579 & 0.7807 & 0.9151 & 0.2472 \\
	0.355 & 0.7119 & 0.1492 & 0.8043 & 0.9247 & 0.2268 \\
	0.360 & 0.7212 & 0.1411 & 0.8251 & 0.9332 & 0.2086 \\
  \end{tabular}
\end{ruledtabular}
\end{table}

\newpage
%

\end{document}